\documentclass[apj]{emulateapj}

\usepackage{subfigure}

\slugcomment{Accepted for publication in the ApJ: January 20, 2009}

\newcommand{\kms}{\textrm{km~s$^{-1}$}}

\newcommand{\cms}{\textrm{cm~s$^{-1}$}}
\newcommand{\ergs}{\textrm{erg~s$^{-1}$}}
\newcommand{\distance}{30~\textrm{Mpc}}
\newcommand{\lsolar}{L$_{\odot}$}
\newcommand{\ml}{M$_{\odot}$ yr$^{-1}$}
\newcommand{\lastday}{935}
\newcommand{\firstday}{7}

\newcommand{\Qo}{$Q_{\lambda}\propto\lambda^{-1}$}

\begin{document}

\title{Near-Infrared Photometry of the Type IIn SN 2005ip:\\ The Case for Dust Condensation}
\shorttitle{Dust Condensation in the Type IIn SN 2005ip}
\author{Ori Fox, Michael F. Skrutskie, Roger A. Chevalier, Srikrishna Kanneganti, Chan Park, John Wilson, Matthew Nelson, Jason Amirhadji, Danielle Crump, Alexi Hoeft, Sydney Provence, Benjamin Sargeant, Joel Sop, Matthew Tea, Steven Thomas, Kyle Woolard}
\affil{Department of Astronomy, University of
Virginia, P.O. Box 400325, Charlottesville, VA 22904}
\email{ofox@virginia.edu}

\begin{abstract}
Near-infrared photometric observations of the Type IIn SN 2005ip in NGC 2906 reveal large fluxes ($>$1.3 mJy) in the $K_s$-band over more than 900 days.  While warm dust can explain the late-time $K_s$-band emission of SN 2005ip, the nature of the dust heating source is ambiguous.  Shock heating of pre-existing dust by post-shocked gas is unlikely because the forward shock is moving too slowly to have traversed the expected dust-free cavity by the time observations first reveal the $K_s$ emission.  While an infrared light echo model correctly predicts a near-infrared luminosity plateau, heating dust to the observed temperatures of $\sim$1400-1600 K at a relatively large distance from the supernova ($\ga 10^{18}$~cm) requires an extraordinarily high early supernova luminosity ($\sim1\times10^{11}$~\lsolar).  The evidence instead favors condensing dust in the cool, dense shell between the forward and reverse shocks.  Both the initial dust temperature and the evolutionary trend towards lower temperatures are consistent with this scenario.  We infer that radiation from the circumstellar interaction heats the dust.  While this paper includes no spectroscopic confirmation, the photometry is comparable to other SNe that do show spectroscopic evidence for dust formation.  Observations of dust formation in SNe are sparse, so these results provide a rare opportunity to consider SNe Type IIn as dust sources.
\end{abstract}


\section{Introduction}
\label{sec_intro}

For nearly 40 years, core-collapse SN events have been considered as possible sources of dust in the universe \citep{cernuschi67,hoyle70}.  More recent studies have proposed that the core-collapse supernovae may be the primary sources of dust in the early universe \citep{todini01, nozawa03, dwek07}.  Several models \citep{todini01,nozawa03,nozawa08} succeed in producing the large amounts of dust observed at high redshifts, and recent data present the first evidence for a supernova origin for dust in an object at $z>6$ \citep{maiolino04}.  

Nonetheless, direct observational evidence for dust formation in supernovae remains sparse, even in the local universe \citep[][and references therein]{meikle07}.  Any newly formed dust would produce a late-time near-infrared excess in comparison to the blackbody optical spectrum.  \citet{merrill80} was the first to detect such an infrared excess from a supernova (SN 1979C).  Since then, only a handful of near-infrared excesses associated with core-collapse events have been observed \citep[e.g.][and references therein]{dwek92,gerardy02,pozzo04,meikle06,smith08a}.  

\begin{figure*}[t]
\plotone{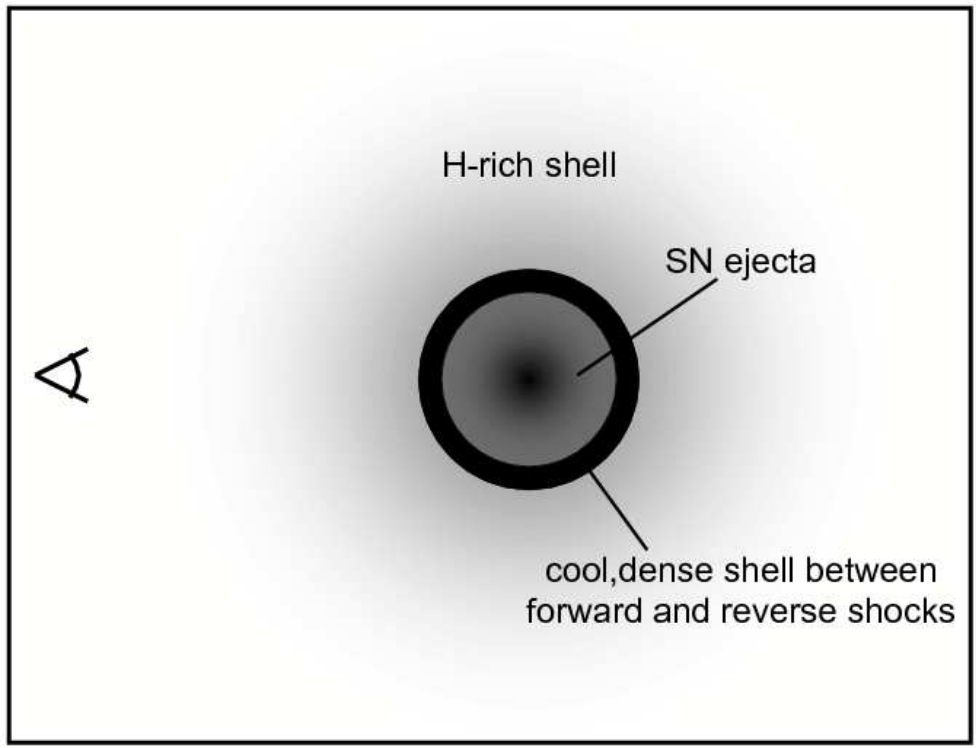}
\caption{An illustration of the environment surrounding core-collapse supernovae similar to SN 2005ip (adapted from \citet{smith08a}).  Dust may form in a number of locations.   Following the initial supernova explosion, the SN ejecta expand and cool sufficiently for dust formation to occur.   A shocked region forms in between a forward shock in the circumstellar gas and a reverse shock in the supernova ejecta.  If the gas is able to cool efficiently, dust condensation is likely to occur in the cool, dense shell behind the shock.  Radiative cooling is more likely at the reverse shock because of the higher density, lower shock velocity, and the possibility of heavy element enrichment.  Cooling at the forward shock is less likely, but the presence of clumps in the circumstellar medium, as indicated by the relatively narrow lines observed in SNe IIn, may allow radiative shocks.}
\label{f1}
\end{figure*}

Late-time near-infrared emission, however, is not unique to dust formation.  Several possible mechanisms may give rise to such an excess due to the complex nature of the surrounding environment of most core-collapse supernovae (see Figure \ref{f1}).  (1) As the ejecta expand and cool, dust condenses and is radiatively heated.  (2) The hot gas behind the forward shock collisionally heats pre-existing circumstellar gas and dust.  (3)  The SN peak luminosity heats pre-existing dust and produces an ``IR echo.''  (4) Dust condenses in either the forward or reverse shocks and is heated by radiative shock emission.

The first scenario requires dust to condense from the ejecta as it expands and cools.  The physical conditions in an expanding SN support this possibility.  In particular, large abundances of the necessary elements, cooling of the ejecta via expansion, and dynamical instabilities can result in dust formation \citep{lucy91,dwek92a,meikle93,roche93}.  In this scenario the primary energy source is the radioactive power released in the expanding gas, which sets the luminosity of the dust.  It is possible, however, for other energy sources to exist, such as circumstellar interactions.

The second scenario describes pre-existing circumstellar dust heated directly by shock interactions.  The scenario is more likely to occur at later times.  Assuming a spherically symmetric circumstellar distribution of dust,  \citet{dwek83b} shows that a SN peak-luminosity of $\sim10^{10}~$\lsolar~creates a dust free cavity, via photo-evaporation, with a radius of $\sim3\times10^{17}$~cm (0.1 light years).  For typical expansion velocities ($\sim5000$~\kms), this cavity delays the onset of emission by a couple of years.  In addition, dust sputtering may limit the grain radiation efficiency so that only a small portion of the total shock power is converted into radiation \citep{draine81}.

In the third scenario, the energy from the SN peak-luminosity once again creates a dust free cavity.  The remaining shell of dust warms to high temperatures but does not vaporize.  Due to light travel time effects, the thermal radiation from the dust grains reaches the observer over an extended period of time \citep{dwek83b}.  The infrared luminosity plateau occurs on year long time scales, which are set by the light travel time across the inner edge of the cavity.  An IR echo's spectrum and temporal evolution are useful for estimating the mass-loss history of the progenitor star, and to investigate the geometry and composition of the circumstellar structure \citep{dwek83b,dwek85,emmering88}.

In the final scenario, the ejecta collide with the pre-existing circumstellar medium, which creates a forward shock.  The circumstellar interaction decelerates the blast wave, thereby simultaneously creating a reverse shock.  The deceleration converts some of the kinetic energy of the shock into X-rays and visible light.  Dust grains are likely to condense in the cool, dense shell that forms behind the radiative shocks as they undergo a thermal instability \citep{pozzo04}.  Typically, the post-shock conditions surrounding reverse shocks are more conducive to dust condensation as the chemically rich, dense ejecta pass through the shock front and cool.  

The decelerated blast wave associated with the reverse shock is more likely to cool in sufficiently dense circumstellar environments.  Type IIn events typically have the densest environments of any SNe, as they are defined by the ``narrow'' H lines that originate from clumps in the pre-existing dense H-rich shell that is moving at a relatively slow velocity \citep{schlegel90}.  Therefore, one might suppose dust formation in the cool, dense shell behind the reverse shock to be associated with Type IIn events.  Unfortunately, Type IIn events are rare, consisting of only $\sim$2-3\% of all core-collapse SNe \citep{galyam07}, and few well-studied events exist.  

This paper presents an analysis of the observed late-time near-infrared emission in the Type IIn SN 2005ip from days \firstday-\lastday~post-discovery.  In \S \ref{sec_observations}, we present the observations, data reduction techniques, and photometry.  In \S \ref{sec_analysis}, we discuss the different possible dust heating mechanisms and compare the SN to SNe with similar characteristics.  The observations suggest that either an IR echo or dust formation in the post-shocked gas explain the long lived IR excess, but that dust formation is more likely.  \S \ref{sec_conclusion} summarizes these results.

\section{Observations}
\label{sec_observations}

\subsection{SN 2005ip}
\label{sec_2005ip}

Supernova 2005ip was discovered in NGC 2906 on UT 2005 November 5 \citep{boles05}.  NGC 2906 lies at a distance of \distance, given a host galaxy recession velocity of 2140 \kms \citep{devaucouleurs91} and $H_{0}~=~72$~\kms~Mpc$^{-1}$ \citep{freedman01}.  Early optical spectra suggested the discovery occurred a few weeks following the explosion \citep{modjaz05}.  The presence of H in the spectrum led to the classification of the SN as a Type II event \citep{modjaz05}.  Late spectra of SN 2005ip showed the development of a narrow H$\alpha$ line that is characteristic of Type IIn supernovae \citep{smith08b}.  Various observers accumulated optical photometry of SN 2005ip over the first 100 days (Teamo 2007\footnote{http://www.astrosurf.com/snweb2/2005/05ip/05ipMeas.htm}).  The supernova was brightest, $V\sim14.8$, at the time of the first observation.  The fact that all data were obtained post-peak does not allow for the explosion date to be constrained.  An unfiltered $R$-band light curve beginning on day $\sim$10 post-discovery from \citet{smith08b} shows a linear magnitude decline through day $\sim$150 post-discovery followed by luminosity plateau at $R\sim17.8$ that begins no later than day $\sim$300 and persists throughout the duration of our observations.  Observations from {\it Swift} on day 461 reveal a magnitude $V\sim18.5$ and a $0.2-10$ keV X-ray luminosity of $1.6\pm0.3\times 10^{40}$ erg s$^{-1}$ \citep{immler07}.

\subsection{Near-Infrared Photometry}
\label{sec_phot}

\begin{figure}
\begin{center}
\epsscale{0.90}
\plotone{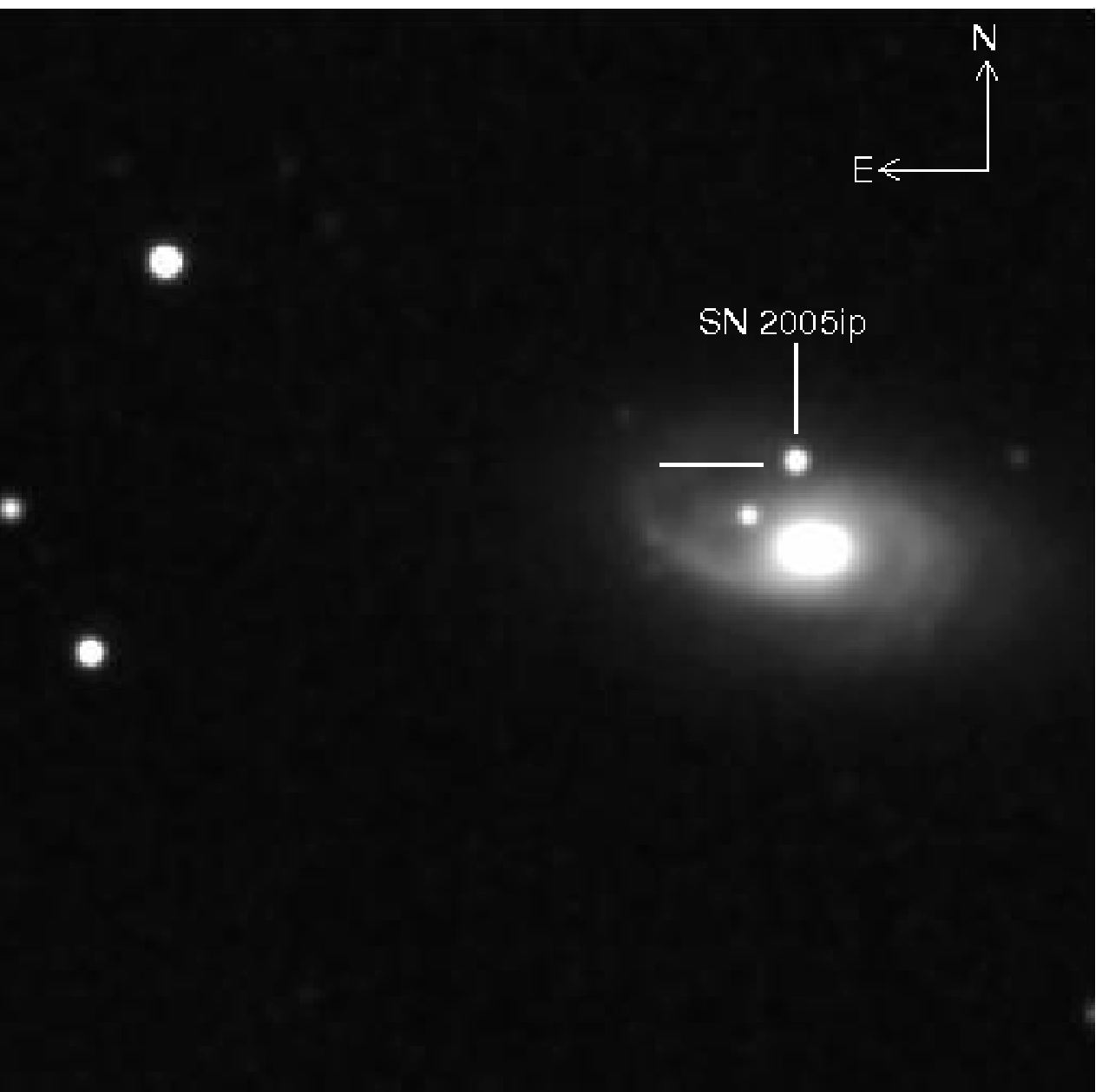}\\
\plotone{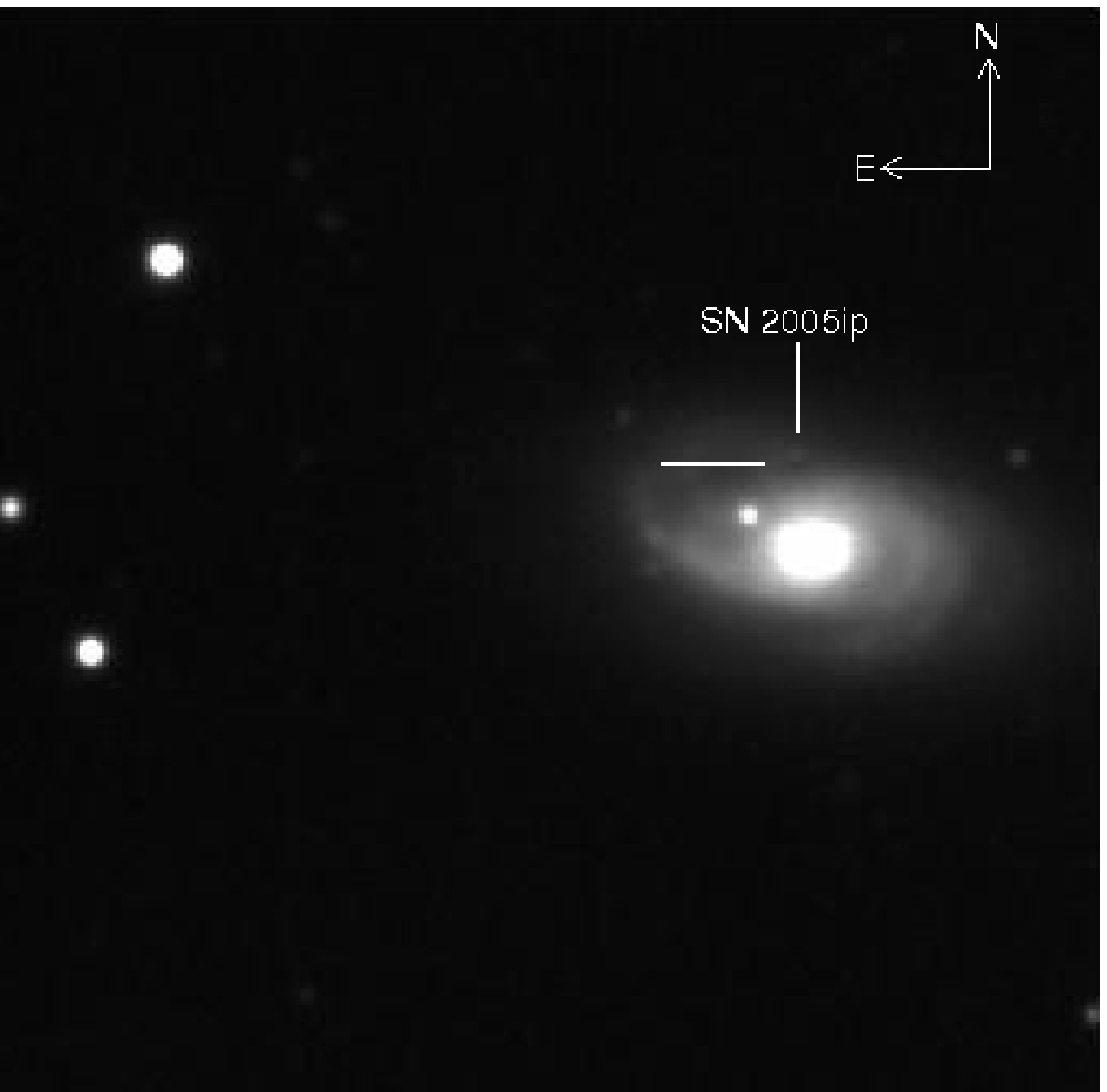}\\
\plotone{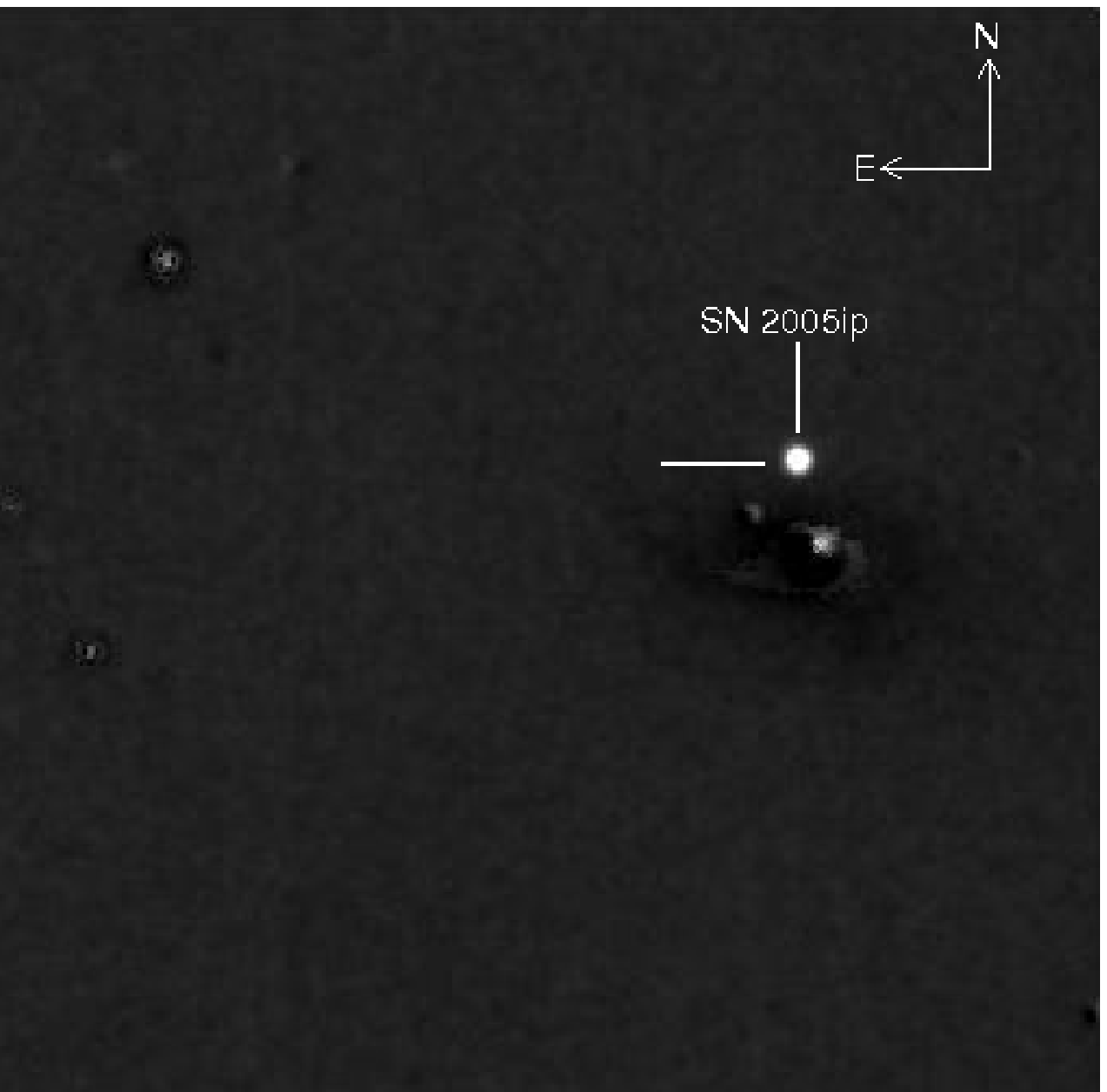}\\
\caption{Figure \ref{f2}a (top) shows a $J$-band image of SN 2005ip within NGC 2906 at \firstday~days post-discovery.  On the left hand side are three 2MASS catalogued stars, which serve as calibration references.  Figure \ref{f2}b (middle) shows the artificial template produced using PSF subtraction, and Figure \ref{f2}c (bottom) shows the original image following template subtraction.  An imperfect subtraction leaves a slight residual, which is enhanced in Figure \ref{f2}c by the contrast settings.  Removing the underlying galaxy via template subtraction minimizes photometric error.\label{f2}}
\epsscale{1}
\end{center}
\end{figure}

Our observations of SN 2005ip consist of $J$-,$H$-, and $K_s$-band images spanning from 2005 December through 2008 June, days \firstday-\lastday~following discovery.  All observations were made with FanCam, a 1024~$\times$~1024 HAWAII-I HgCdTe imaging system on the University of Virginia's 31-inch telescope at Fan Mountain, just outside of Charlottesville, VA \citep{kanneganti07}.  Each epoch consists of fifteen minutes of integration in $JHK_s$~bands, which have detection limits at the 10$\sigma$~level of 0.066, 0.098, and 0.156 mJy (or 18.5, 17.5, and 16.5 mag), respectively.  Individual exposures are sky background limited and have an integration time of either 30 or 60 seconds.  Flat-field frames are composed of dusk and dawn sky observations.  We employed standard near-infrared data reduction techniques in IRAF.  The relatively small galaxy size made it possible to fit it into a single array quadrant.  Empty quadrants were efficiently utilized as sky exposures.  Data were taken with the galaxy placed in each quadrant and each quadrant was reduced separately.  Ultimately, all reduced quadrants were coadded.  Figure \ref{f2}a shows a typical reduced $J$-band image.

Template subtraction minimizes photometric confusion from the underlying galaxy.  While there are no existing FanCam observations of the galaxy prior to SN 2005ip, IRAF's DAOPHOT PSF and SUBSTAR packages provide the means to create and subtract an average PSF at the SN location for each epoch (Figure \ref{f2}b).  A coaddition of each epoch yields a master template that is deeper and smoother than any individual epoch.  The subtraction of the template from each epoch leaves only the SN (Figure \ref{f2}c).  To ensure a clean subtraction, individual and template PSFs are scaled to match in shape and intensity using IRAF's PSFMATCH package.  We performed photometry with IRAF's APPHOT package, implementing an 8-pixel radius aperture and a 3-pixel wide annulus with an 8-pixel inner radius.  For magnitude calibration, the SN is compared to three 2MASS reference stars located in the eastern part of Figure \ref{f2}.

\begin{deluxetable*}{ c c c c c c c c }[t]
\tablewidth{0pt}
\tabletypesize{\tiny}
\tablecaption{Near-Infrared Photometry of SN 2005ip \label{tab1}}
\tablecolumns{8}
\tablehead{
\colhead{JD} & \colhead{Epoch} & \colhead{J} & \colhead{H} &
\colhead{K$_{\rm s}$} & \colhead{J} & \colhead{H} & \colhead{K$_{\rm s}$}\\
\colhead{(2450000+)} & \colhead{(days)} & \colhead{(mag)} & \colhead{(mag)} & \colhead{(mag)} & \colhead{(mJy)} & \colhead{(mJy)} & \colhead{(mJy)}
}
\startdata
3686.5 & 7.0 & 14.42(0.01) & - & 13.87(0.01) & 2.84(0.03) & - & 1.75(0.02) \\
3688.0 & 8.0 & 14.41(0.01) & - & 13.93(0.02) & 2.88(0.03) & - & 1.66(0.03) \\
3743.0 & 63.0 & 15.42(0.03) & 14.91(0.05) & 14.12(0.03) & 1.14(0.03) &
1.07(0.05) & 1.39(0.03) \\
3763.1 & 83.0 & 15.51(0.03) & 14.71(0.03) & 13.82(0.02) & 1.04(0.03) &
1.28(0.04) & 1.83(0.03) \\
3773.0 & 93.0 & 15.59(0.03) & 14.66(0.03) & 13.75(0.01) & 0.97(0.02) &
1.34(0.04) & 1.96(0.03) \\
3799.5 & 120.0 & 15.84(0.03) & 14.66(0.03) & 13.67(0.01) & 0.77(0.02) &
1.34(0.04) & 2.11(0.03) \\
3830.7 & 151.0 & 16.08(0.04) & 14.77(0.03) & 13.72(0.01) & 0.62(0.02) &
1.21(0.04) & 2.01(0.03) \\
3850.5 & 171.0 & 16.21(0.05) & 14.81(0.04) & 13.75(0.02) & 0.54(0.02) &
1.16(0.04) & 1.97(0.04) \\
3865.5 & 186.0 & 15.96(0.04) & 14.75(0.04) & 13.67(0.01) & 0.69(0.02) &
1.23(0.04) & 2.12(0.03) \\
3879.5 & 200.0 & 16.05(0.05) & 14.78(0.04) & 13.67(0.02) & 0.63(0.03) &
1.20(0.04) & 2.10(0.03) \\
4030.9 & 351.0 & 16.25(0.05) & - & 13.75(0.01) & 0.53(0.02) & - & 1.96(0.03) \\
4112.0 & 432.0 & 16.49(0.06) & 15.20(0.05) & 13.87(0.01) & 0.42(0.02) &
0.82(0.04) & 1.75(0.02) \\
4159.8 & 480.0 & 16.60(0.07) & 15.15(0.05) & 13.89(0.02) & 0.38(0.02) &
0.85(0.04) & 1.73(0.03) \\
4227.6 & 548.0 & 16.66(0.08) & 15.23(0.06) & 13.96(0.02) & 0.36(0.03) &
0.79(0.04) & 1.62(0.03) \\
4367.9 & 688.0 & 16.76(0.08) & 15.48(0.07) & 13.92(0.02) & 0.33(0.02) &
0.63(0.04) & 1.68(0.03) \\
4406.8 & 722.0 & 16.85(0.09) & 15.52(0.07) & 14.03(0.02) & 0.30(0.03) &
0.61(0.04) & 1.52(0.03) \\
4452.9 & 767.0 & 16.94(0.11) & 15.67(0.08) & 14.07(0.02) & 0.28(0.03) &
0.53(0.04) & 1.46(0.02) \\
4516.8 & 831.0 & - & 15.87(0.11) & 14.14(0.03) & - & 0.44(0.05) & 1.37(0.04) \\
4537.6 & 852.0 & 17.03(0.10) & 15.80(0.09) & 14.10(0.02) & 0.26(0.02) &
0.47(0.04) & 1.42(0.03) \\
4580.6 & 895.0 & 17.01(0.10) & 15.83(0.09) & 14.12(0.02) & 0.26(0.02) &
0.45(0.04) & 1.40(0.02) \\
4620.6 & 935.0 & 17.17(0.12) & 16.14(0.15) & 14.18(0.05) & 0.23(0.03) &
0.34(0.05) & 1.32(0.06)
\enddata
\end{deluxetable*}

\begin{figure*}
\begin{center}
\plotone{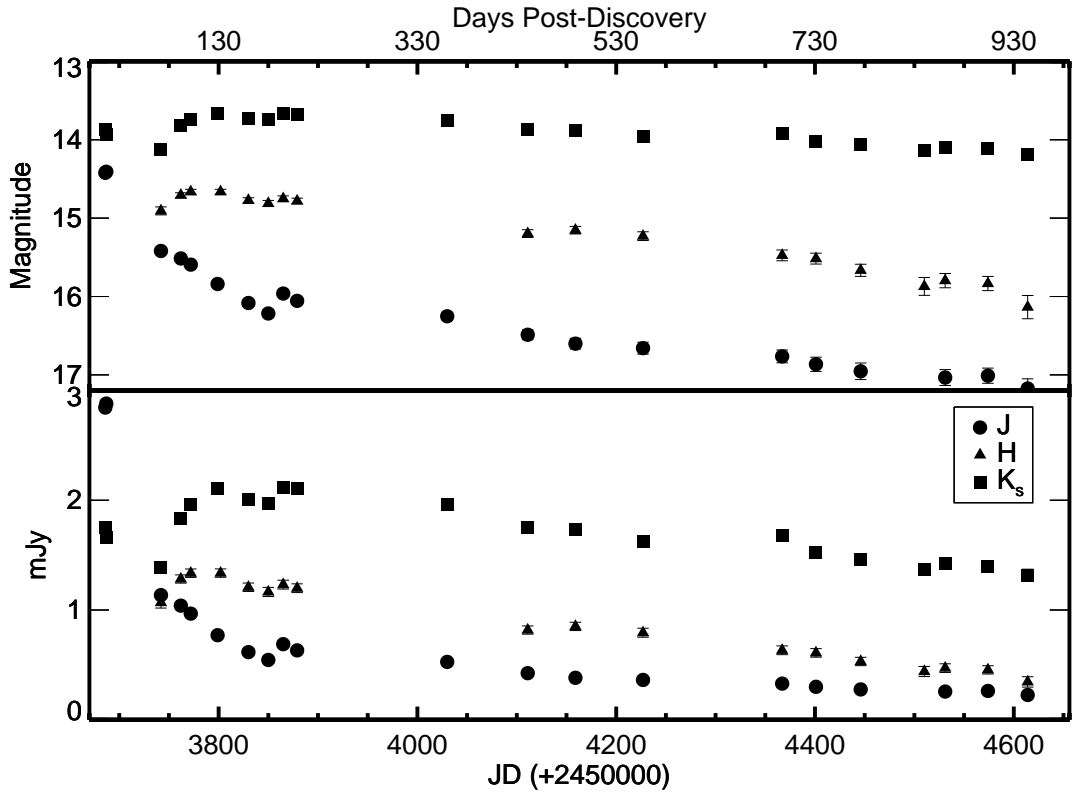}
\caption{$JHK_s$ light curves for SN 2005ip showing magnitude (top) and flux (bottom).  The $K$$_s$-band curve remains fairly constant until late-times, even as the $J$-band declines.  The observed kink in the $H$~and $K$~ light curves at day $\sim$50 is likely due to the onset of dust formation in the post-shocked gas.}
\label{f3}
\end{center}
\end{figure*}

\begin{deluxetable}{ c c c c }
\tablewidth{0pt}
\tabletypesize{\tiny}
\tablecaption{Colors of SN 2005ip \label{tab2}}
\tablecolumns{4}
\tablehead{
\colhead{Epoch} & \colhead{J-H} & \colhead{J-K$_{\rm s}$} & \colhead{H-K$_{\rm s}$}\\
\colhead{(days)} & \colhead{(mag)} & \colhead{(mag)} & \colhead{(mag)}
}
\startdata
7 & - & 0.55(0.02) & - \\
8 & - & 0.48(0.02) & - \\
63 & 0.51(0.06) & 1.30(0.04) & 0.78(0.06) \\
83 & 0.81(0.04) & 1.69(0.03) & 0.89(0.04) \\
93 & 0.93(0.04) & 1.85(0.03) & 0.92(0.03) \\
120 & 1.18(0.05) & 2.17(0.04) & 0.99(0.03) \\
151 & 1.31(0.06) & 2.36(0.05) & 1.05(0.04) \\
171 & 1.40(0.06) & 2.47(0.05) & 1.07(0.04) \\
186 & 1.21(0.05) & 2.30(0.04) & 1.09(0.04) \\
200 & 1.27(0.06) & 2.38(0.05) & 1.11(0.04) \\
351 & - & 2.50(0.05) & - \\
432 & 1.29(0.08) & 2.62(0.06) & 1.33(0.05) \\
480 & 1.45(0.08) & 2.72(0.07) & 1.27(0.05) \\
548 & 1.43(0.09) & 2.70(0.08) & 1.27(0.06) \\
688 & 1.29(0.11) & 2.85(0.08) & 1.56(0.07) \\
722 & 1.35(0.11) & 2.84(0.09) & 1.49(0.07) \\
767 & 1.29(0.13) & 2.89(0.11) & 1.60(0.08) \\
831 & -15.87(0.11) & -14.14(0.03) & 1.73(0.12) \\
852 & 1.24(0.14) & 2.94(0.10) & 1.70(0.09) \\
895 & 1.18(0.13) & 2.90(0.10) & 1.72(0.09) \\
935 & 1.04(0.19) & 2.99(0.13) & 1.95(0.16)
\enddata
\end{deluxetable}

This template subtraction method yields results similar to, but less noisy than, other techniques, such as aperture photometry and PSF photometry without template subtraction.  Table \ref{tab1} lists the $JHK_s$ photometry and Figure \ref{f3} plots the light curve.  The $K_s$-band magnitude remains fairly constant at $\leq$ 14.2 mag (a flux of $\geq$ 1.3 mJy) for the duration of the observations, even as the $J$-band declines. A notable `kink' occurs at day $\sim$200, although we do not speculate on the cause in this paper.  Table \ref{tab2} lists the near-infrared color evolution.  Figure \ref{f4} shows the color to quickly redden between $\sim$50 and 200 days.  This trend becomes more gradual and continues throughout the last observation epoch.

\section{Analysis and Discussion}
\label{sec_analysis}

\subsection{Temperature and Luminosity Evolution}
\label{sec_temp}

Figure \ref{f5} shows the near-infrared fluxes.  Matching the $H-K_s$ colors to the Planck function, presuming a dust emission/absorption efficiency, $Q(\lambda)$, establishes a near-infrared dust temperature.  The flux of a single dust particle of radius $a$ at a temperature $T_d$ and wavelength $\lambda$ is given by
\begin{equation}
F_{\lambda,T}=Q(\lambda)B_\lambda(T),
\label{eqn:blackbody}
\end{equation}
where $B_\lambda(T)$ is the Planck function.  The emission/absorption efficiency is given as
\begin{eqnarray}
Q = \left(\frac{a}{\lambda}\right)^n,~{\rm for}~a\ll\lambda\\
Q = 1, ~{\rm for}~a\gg\lambda
\label{eqn:emissivity}
\end{eqnarray}
where $n$ is an integer.  For large grain sizes (e.g. $a\geq1$~\micron), $n=0$ because the dust acts as a blackbody.  For smaller grains (i.e. $a\ll1\micron$), $n>0$.  We consider the cases $n$=0, 1, and 2 to account for both large grains that are associated with condensing dust \citep{pozzo04} and smaller grains that are associated with pre-existing, optically thin dust, which is characterized by $\lambda^{-1}$ and $\lambda^{-2}$ emission/absorption efficiencies \citep{dwek83a}.  

\begin{figure}
\begin{center}
\plotone{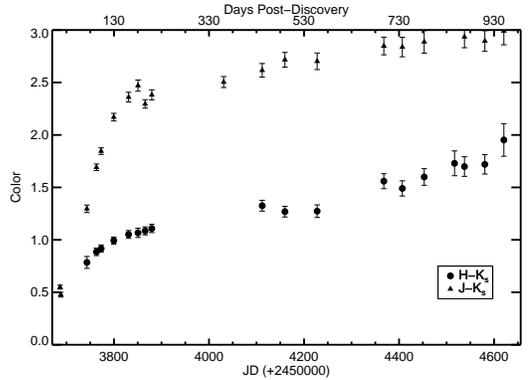}
\caption{The $J-K_s$ and $H-K_s$ color evolution for SN 2005ip.  The $J-K_s$ color reveals the quickly declining photospheric component in the $J$-band at early times.  The colors continue to redden throughout the observations as the warm dust dominates the infrared emission.}
\label{f4}
\end{center}
\end{figure}

Figure \ref{f5} plots the corresponding Planck curves over the original fluxes.  The fits exclude the $J$-band fluxes because of SN photospheric flux and line emission contamination, especially at early times when the photospheric temperature is $\sim$10,000 K.  At these early times, the $H$-band flux may also contain some contribution from the SN photosphere.  By later times, however, we presume the photosphere has faded sufficiently so that the warm dust dominates the flux.  With only 2 data points for each epoch there is no clear `best' fit, so it is necessary to consider all cases.

Overall, the Planck peaks trend towards longer wavelengths/cooler temperatures.  Table \ref{tab3} lists the $H-K_s$ color temperatures for each emission/absorption efficiency at each epoch.  Figure \ref{f6}a plots the color temperature evolution.  In all cases, the fading SN photosphere contributes to the dramatic temperature decline over the first $\sim$150 days.  Although the exact point at which the dust temperature begins to dominate is unknown, we presume the photospheric component to decline sufficiently below the dust temperature after $\sim$150 days, at which point the temperature begins a slow, steady decline.  While the chosen efficiency can alter the derived dust temperatures at early times by a few hundred Kelvin, these temperatures are less than or equal to the dust vaporization temperature ($\sim1400-1600$ K).  Over the next 2 years, the temperatures drop to $\sim$1000 K.  Considering the observations are made at near-infrared wavelengths, they probe the warmest and most luminous grains.  Cooler dust may also be present, but its emission peaks at wavelengths longer than those observed here.

In contrast to the decreasing dust temperature, the dust luminosity plateaus early on and does not decline throughout the extent of the observations.  Assuming a distance of \distance, integrating the Planck curves over all wavelengths yields a lower limit to the bolometric dust luminosity, shown in Figure \ref{f6}b and listed in Table \ref{tab3}.  Again, this calculation does not account for cooler dust that could contribute at longer wavelengths.  Throughout the entire set of observations, the dust luminosity remains at $\sim3~\times~10^{41}$~\ergs.

\begin{deluxetable*}{ c c c c c c c c }
\tablewidth{0pt}
\tabletypesize{\tiny}
\tablecaption{Temperatures and Luminosities of SN 2005ip \label{tab3}}
\tablecolumns{8}
\tablehead{
\colhead{Epoch} & \colhead{T$_{bb}$} & \colhead{T$_{\lambda^{-1}}$} &
\colhead{T$_{\lambda^{-2}}$} & \colhead{L$_{bb}$} & \colhead{L$_{\lambda^{-1}}$}
& \colhead{L$_{\lambda^{-2}}$} & \colhead{L$_{\rm radioactive}$} \\
\colhead{(d)} & \colhead{(K)} & \colhead{(K)} & \colhead{(K)} &
\colhead{$(10^{41} \rm erg~s^{-1})$} & \colhead{$(10^{41} \rm erg~s^{-1})$} &
\colhead{$(10^{41} \rm erg~s^{-1})$} & \colhead{$(10^{41} \rm erg~s^{-1})$}
}
\startdata
7 & - & - & - & - & - & - & 42.09 \\
8 & - & - & - & - & - & - & 42.09 \\
63 & 1969(89) & 1547(56) & 1279(39) & 41.43(0.03) & 41.38(0.03) & 41.34(0.03) &
41.86 \\
83 & 1809(47) & 1449(31) & 1213(22) & 41.53(0.01) & 41.48(0.02) & 41.43(0.02) &
41.80 \\
93 & 1765(44) & 1422(29) & 1194(20) & 41.57(0.01) & 41.50(0.01) & 41.46(0.01) &
41.75 \\
120 & 1667(38) & 1359(26) & 1149(18) & 41.59(0.01) & 41.53(0.01) & 41.49(0.01) &
41.64 \\
151 & 1599(39) & 1314(27) & 1118(19) & 41.57(0.01) & 41.50(0.01) & 41.46(0.01) &
41.53 \\
171 & 1582(45) & 1303(31) & 1110(22) & 41.56(0.02) & 41.50(0.02) & 41.46(0.02) &
41.45 \\
186 & 1561(38) & 1289(26) & 1100(19) & 41.60(0.01) & 41.53(0.01) & 41.49(0.01) &
41.39 \\
200 & 1538(38) & 1273(26) & 1088(19) & 41.59(0.01) & 41.53(0.01) & 41.49(0.02) &
41.34 \\
351 & - & - & - & - & - & - & 40.75 \\
432 & 1345(39) & 1140(28) & 990(21) & 41.53(0.02) & 41.48(0.02) & 41.42(0.02) &
40.43 \\
480 & 1390(41) & 1172(29) & 1013(22) & 41.53(0.02) & 41.46(0.02) & 41.41(0.02) &
40.25 \\
548 & 1386(48) & 1169(34) & 1012(26) & 41.50(0.02) & 41.43(0.02) & 41.38(0.02) &
39.98 \\
688 & 1186(42) & 1024(31) & 902(24) & 41.59(0.02) & 41.50(0.02) & 41.43(0.02) &
39.43 \\
722 & 1229(45) & 1056(34) & 926(26) & 41.52(0.02) & 41.43(0.02) & 41.38(0.02) &
39.31 \\
767 & 1163(45) & 1007(33) & 888(26) & 41.53(0.02) & 41.45(0.02) & 41.39(0.02) &
39.14 \\
831 & 1093(59) & 954(45) & 847(35) & 41.55(0.03) & 41.46(0.03) & 41.39(0.04) &
38.89 \\
852 & 1108(48) & 966(36) & 856(28) & 41.56(0.03) & 41.46(0.03) & 41.39(0.03) &
38.82 \\
895 & 1098(46) & 958(35) & 850(28) & 41.56(0.02) & 41.46(0.03) & 41.39(0.03) &
38.67 \\
935 & 991(64) & 876(50) & 785(40) & 41.63(0.04) & 41.52(0.04) & 41.43(0.04) &
38.52
\enddata
\end{deluxetable*}

\subsection{Comparison to Previous SNe}
\label{sec_comparison}

Comparing SN 2005ip photometry to prior, well-studied events can help expose the dust heating mechanism.  We find three well-sampled SNe with similar bright near-infrared late-time emission: SN 1998S \citep{fassia00,pozzo04}, SN 1995N \citep{gerardy02}, and SN 2006jc \citep{sakon07,smith08a,mattila08,nozawa08,dicarlo08}.  The associated spectra of these SNe reveal further details about the surrounding environment and the nature of the infrared emission.

Type IIn SN 1998S is most comparable to SN 2005ip, with similar luminosities and temperatures, although a decline in the $K$-band luminosity is observed as early as day 333 \citep{fassia00,pooley02,pozzo04}.  \citet{pozzo04} conclude dust condensation in the cool, dense shell behind the reverse shock is responsible for the late-time luminosity.  They do not consider the forward shock given the increased likelihood for dust condensation in the post-shocked ejecta.  All observations are consistent with this scenario.  The observed optical line profiles likely arise from the low-ionization gas in the cooling shock \citep{chevalier94}.  The coincidence of the line velocities with the velocity of the dust shell (derived from the estimated shell radius from a blackbody approximation versus the time since the explosion) suggests a physical relationship between the two.  The condensing dust also explains the fading central and redshifted H$\alpha$ peaks with respect to the blueshifted peak observed by \citet{gerardy02}.  \citet{fassia01} observe CO emission to develop at the same time the absorption is observed.  Finally, the pure blackbody fits of the dust ($n$=0) are more suggestive of dust condensing in optically thick lumps than pre-existing, optically thin dust.

\begin{figure*}
\begin{center}
\plottwo{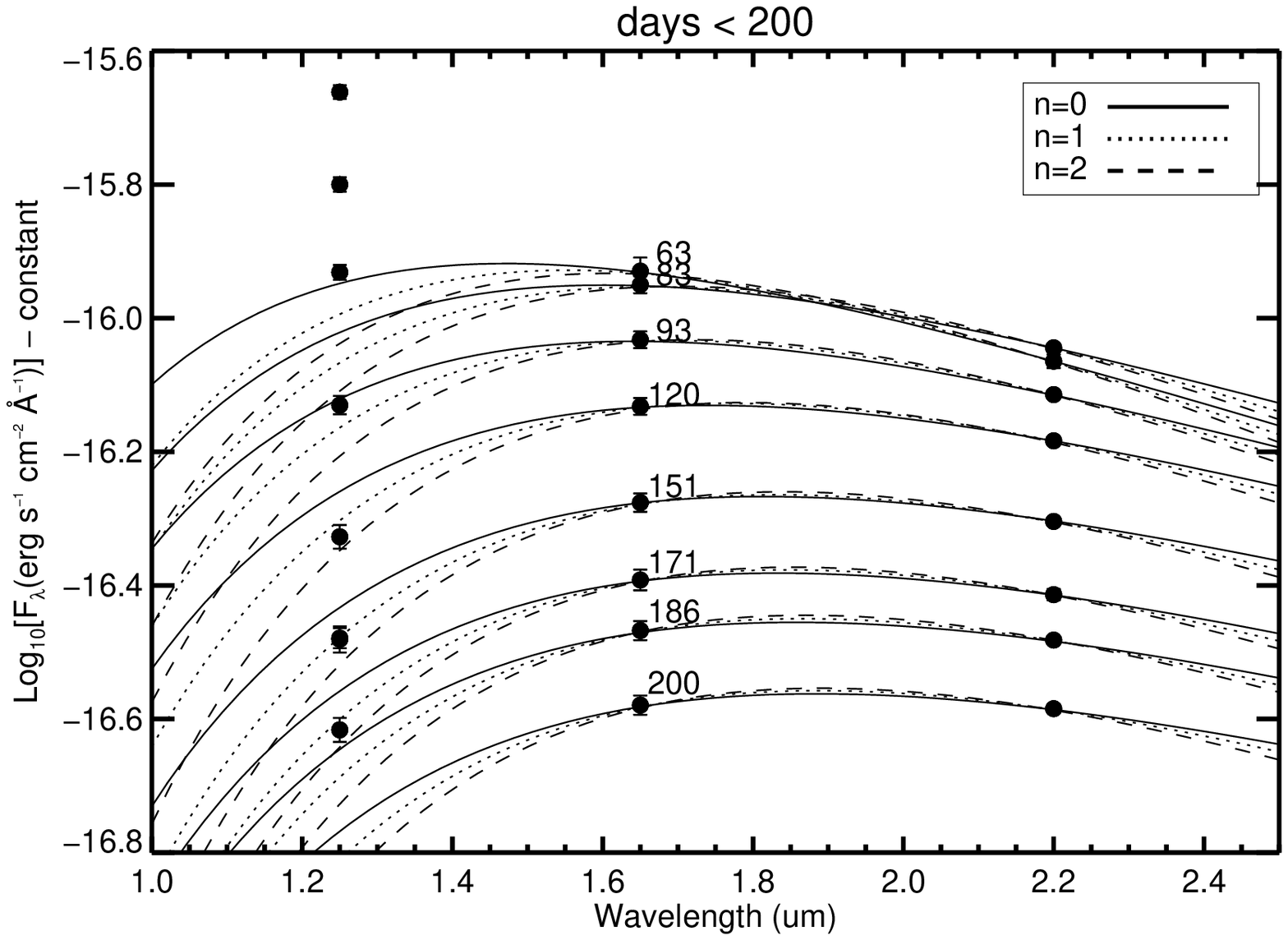}{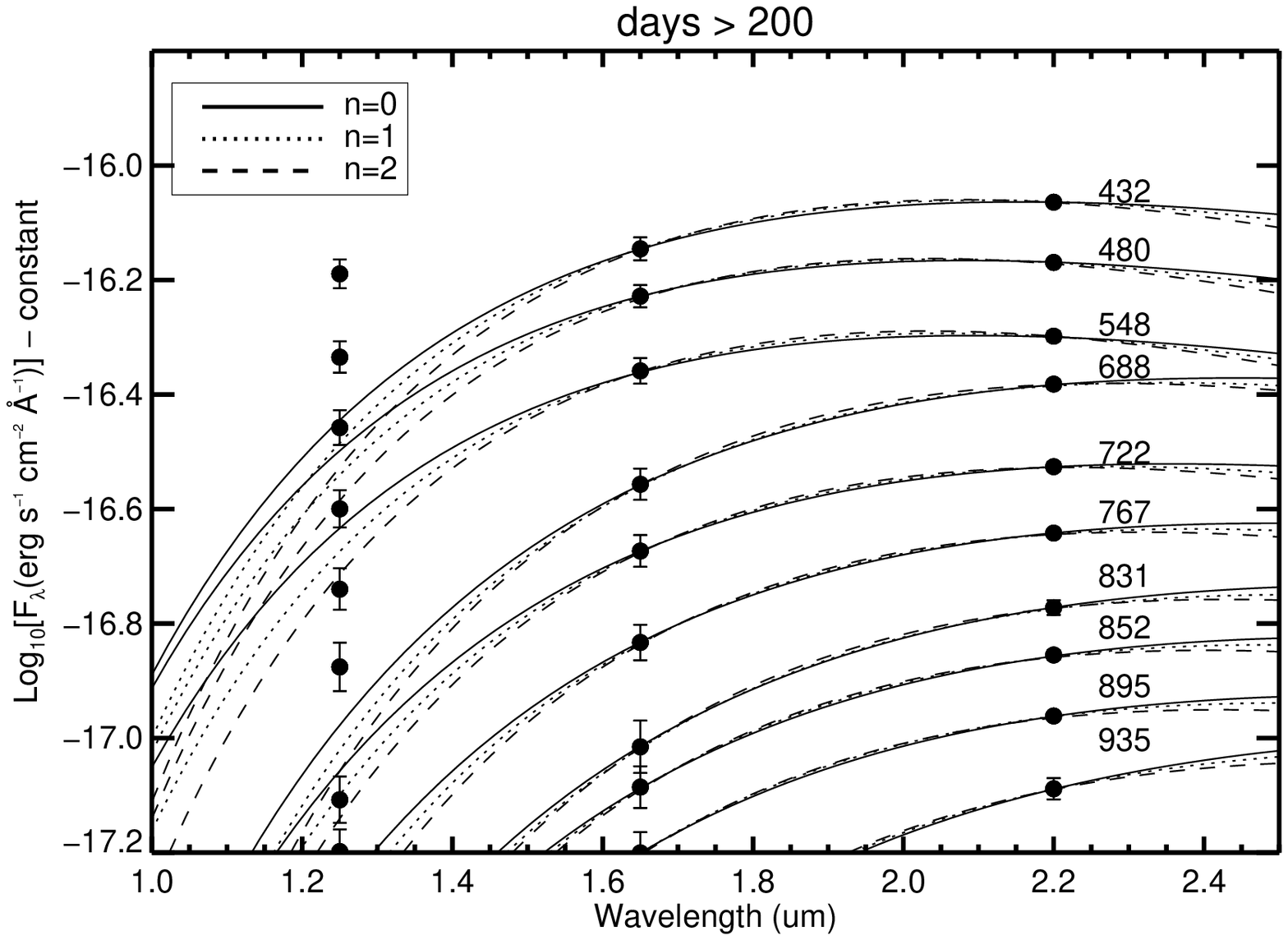}
\caption{Near-infrared fluxes for SN 2005ip at (a) early (top) and (b) late (bottom) times.  Presuming a dust emission/absorption efficiency $Q \propto \frac{1}{\lambda^n}$, where $n$ is an integer, a Planck curve is fit to the $H-K_s$ colors to establish a near-infrared dust temperature.  We consider both large grains that are associated with condensing dust ($n=0$) and smaller grains that are associated with pre-existing optically thin dust ($n=1~\&~2$).  The Planck curves are plotted over the original fluxes.  The curves exclude the $J$-band fluxes because of SN photospheric flux and line emission contamination, especially at early times when the photospheric temperature is $\sim$10,000 K.  These fluxes consistently fall above the Planck curves.  Overall, the Planck peaks trend towards longer wavelengths/cooler temperatures.\label{f5}}
\epsscale{1}
\end{center}
\end{figure*}

Type IIn SN 1995N also shows comparable temperatures and luminosities to SN 2005ip \citep{fox00,gerardy02,pastorello05}.  Similar to SN 2005ip, the luminosity plateau extends all the way through day $\sim1000$, after which a decline was observed out to day 2413 \citep{gerardy02}.  \citet{gerardy02} do not consider dust condensation, but instead suggest the dust is heated by either an IR echo or by the X-rays produced in direct circumstellar interaction heating of pre-existing dust.  While the IR echo model adequately describes the photometric evolution, the implied dust cavity radius, created via heating and vaporization by the peak SN luminosity, requires an unusually bright peak luminosity.  Instead, the authors suggest an event prior to the core collapse could have cleared out the large dust cavity required in this instance.  Alternatively, the authors' circumstellar medium interaction model does not require such a large dust cavity.  The pre-existing dust is heated by X-ray and UV emissions produced by the much more slowly evolving shock interactions with a dense circumstellar medium.

Interestingly, \citet{gerardy02} rule out direct collisional heating of pre-existing dust by the hot, post-shocked electrons.  The authors argue dust temperatures of $800-1000$ K require prohibitively high gas densities ($n_g \ga 10^{21}$ cm$^{-3}$ for $T_e=10^7$ K).  However, we find these densities are not required in this scenario.  For $a$ = 0.1 \micron~silicate grains and a gas temperature $T_e=10^7$~K, \citet{dwek08} show in their figure 1 that for dust temperatures of $\sim$800 K, gas densities are only $n_g\sim10^5~{\rm cm^{-3}}$.  These densities are not prohibitively high and are, in fact, close to what would be expected for typical progenitor mass loss rates ($\sim10^{-4}-10^{-5}$ \ml) and wind velocities ($\sim10~\kms$) at radii of $\sim10^{17}$~cm.

The authors also argue that collisional heating does not predict the observed drop in the infrared luminosity before its X-ray counterpart.  We find, however, that the sputtering lifetime, $\tau_{\rm sput}$, of the dust grains is short enough that there should not necessarily be a correlation between the infrared and X-ray luminosities.  The sputtering lifetime at gas temperatures above $\sim$10$^6$~K given by \citep{dwek92} is
\begin{equation}
\tau_{\rm sput} ({\rm yr}) \approx 1\times10^6 \frac{a(\micron)}{n_g({\rm cm^{-3}})}.
\label{sputter}
\end{equation}
For the grain size ($a$ = 0.1 \micron) and gas densities ($n_g\sim10^5~{\rm cm^{-3}}$) above, $\tau_{\rm sput}\approx1$~year, which is shorter than the observed light plateau in SN 1995N.  In such a short period of time, infrared emission from collisionally heated dust will decrease as the larger dust grains are destroyed.  These effects could explain the lack of correlation between the infrared and X-ray luminosities.  Nonetheless, collisional heating is still unlikely in this case. \citet{draine81} shows that collisionally heated grains radiate only $\sim12\%$ of the shock power before the grains are sputtered, thereby limiting the contribution of infrared emission from collisionally heated dust.

SN 2006jc also shows an extended period of increased late-time infrared emission.  Unlike the previous two examples, SN 2006jc is a Type Ib event with a dense circumstellar medium that is likely associated with a LBV-like outburst originating from the progenitor at -730 days \citep{nakano06,pastorello07}.  Between days 51 and 75, the supernova undergoes a brightening at near-infrared wavelengths \citep{arkharov06,dicarlo08}.  On day 75, \citet{smith08a} measured a luminosity $L\sim 2.5\times 10^{41}~\ergs$ and temperature $T\sim1600$ K.  The temperature stayed constant for only $\sim$1 month and then cooled.  Throughout this time period, X-rays indicate interaction of the supernova shock with the circumstellar shell created by the previous LBV outburst \citep{immler08}.  The SN continued to remain bright at infrared wavelengths through day 493, which was the last observation epoch \citep{mattila08,dicarlo08}.

\begin{figure*}
\begin{center}
\plottwo{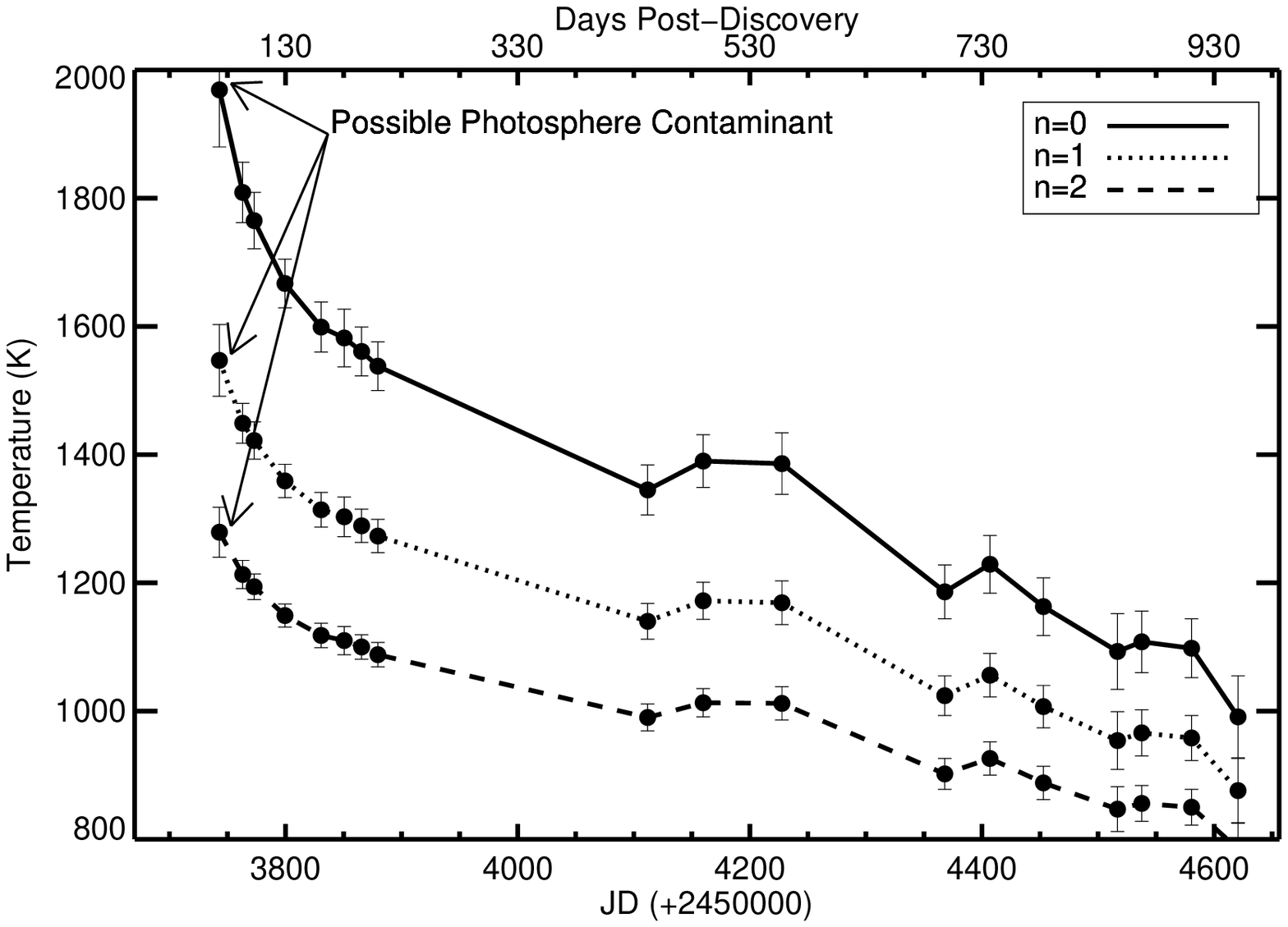}{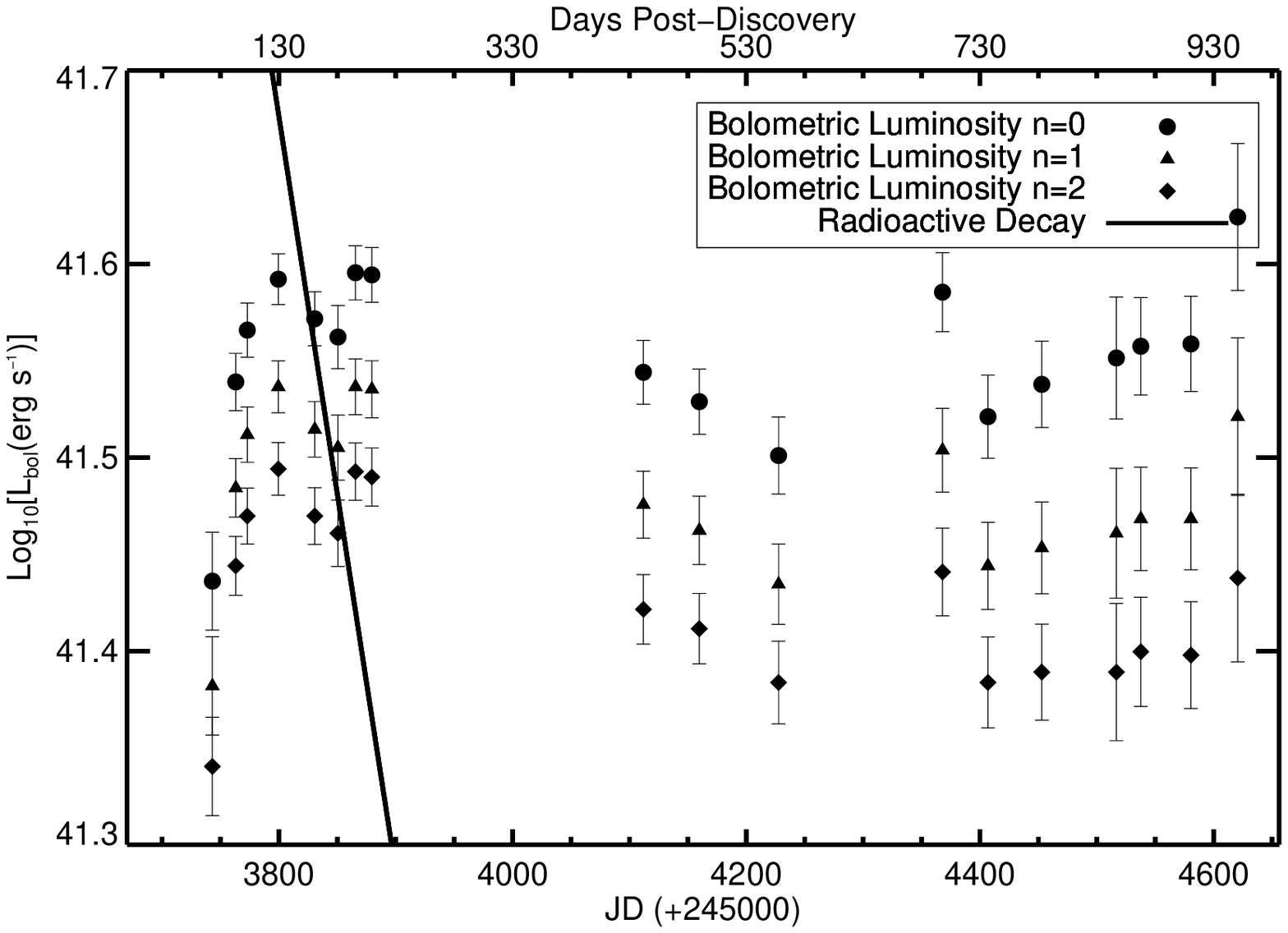}
\caption{The (a) temperature (top) and (b) luminosity (bottom) evolution for SN 2005ip for all dust emission/absorption efficiencies.  At early times ($<$150 days), the derived temperature drops as contamination from the SN photosphere fades.  At later times ($>$150 days) the dust emission dominates.  Although the exact point at which the dust temperature begins to dominate is unknown, we presume the photospheric component to decline sufficiently below the dust temperature after $\sim$150 days.  While the early dust temperatures vary with emission/absorption efficiency by a couple hundred Kelvin, they are all less than or equal to the dust vaporization temperature ($\sim1400-1600$ K).  These near-infrared observations probe only the warmest and most luminous grains.  Cooler dust may also be present, but its emission peaks at longer wavelengths.  Assuming a distance of \distance, integrating the Planck curves over all wavelengths yields an approximate bolometric dust luminosity.  In contrast to the decreasing dust temperature, the dust luminosity plateaus from early times.  It is unlikely that radioactive decay within the ejecta (solid line) is the energy source as the decay curve does not match the observed luminosities.\label{f6}}
\epsscale{1}
\end{center}
\end{figure*}

While warm dust is the widely accepted source of the late-time near-infrared emission, the origin of the dust is debated.  \citet{smith08a} argue the dust is newly formed based on (1) the development of a red/IR continuum consistent with the temperature of newly formed graphite grains or slightly hotter silicate grains, (2) the coincidental fading of the redshifted wings of the narrow He I emission lines, and (3) a simultaneous decline in the expected optical flux \citep{foley07} and brightening in near-infrared.  \citet{sakon07} and \citet{nozawa08} propose the dust condenses in the freely expanding ejecta.  Both \citet{smith08a} and \citet{mattila08} reject this scenario because it cannot explain the observed spectral line reddening.  Instead, these authors favor dust condensation in the cool, dense shell created as the supernova shock interacts with the circumstellar medium.  \citet{smith08a} favor dust condensation in the reverse shock because of the observed carbon-rich features that are consistent with enriched ejecta encountering the reverse shock.  In contrast, \citet{mattila08} conclude the reverse shock would not be radiative and, therefore, not conducive to dust formation.  Instead, the authors argue for dust condensation in the cool, dense shell behind the forward shock.

In any case, the temperature remains constant for about one month at early times.  \citet{smith08a} argue that if there is a sufficient amount of circumstellar material, then grains will continue to form at the vaporization temperature.  To produce the observed luminosity of $L=(2-3)\times 10^{41}$~\ergs~by cooling through $\sim$1600 K, the particle condensation rate is given as  $\dot N=L/\epsilon$, where $\epsilon$ is the thermal energy per particle.  For dust grains, the thermal energy per particle is given as $\epsilon = C \Delta T$, where $C$ is the dust heat capacity \citep{draine01} and $\Delta T$ is 1600 if we assume the dust cools immediately and contributes only to the instantaneous luminosity.  In this case, we calculate that the dust must condense on the order of $\sim$1000 M$_{\odot}$~day$^{-1}$.  This unrealistic mass condensation rate indicates that there must be an additional heating source to power the observed infrared luminosity.  Given the optical light curve, which is powered by radioactivity, quickly declines \citep{foley07} at about the same time as \citet{arkharov06} observe near-infrared brightening suggests that the newly formed dust is absorbing the optical emission from the central supernova.  In this case, the likely dust heating source is the photospheric emission from radioactivity.  \citet{mattila08} similarly argue that the cooling of the newly formed grains is insufficient to account for the observed near-infrared luminosity and conclude a similar heating source for their dust condensation scenario.

\subsection{The Source of the Infrared Excess in SN 2005ip}
\label{sec_excess}

Warm dust surrounding SN 2005ip is the likely source of the near-infrared excess.  As discussed in \S \ref{sec_intro}, there are several possible explanations for the large late-time $K_s$-band luminosity: dust condensation in the ejecta, collisional heating in circumstellar medium shock interaction, an infrared echo, and dust condensation in the cool, dense shell behind either the forward or reverse shocks.  This section briefly explores each possibility in order to distinguish between the different scenarios.  SNe 1998S, 1995N, and 2006jc provide important points of reference.

\subsubsection{Dust Condensation in the Unshocked Ejecta}
\label{sec_ejecta}

Dust condensation may occur in the SN ejecta as it expands and cools.  The available power in the expanding gas, primarily from radioactive decay, dictates the temperature and luminosity evolution of the grains.  The decay of $^{56}$Co, with a half-life of 88 days, dominates the plausible radioactive power sources over the first hundreds of days.  Other radioactive sources with longer half-lives become dominant at later times.  \citet{woosley89} describe the details of radioactive energetics in supernovae.  For SN 2005ip, the radioactive luminosity is larger than the observed dust luminosities at early times, but later falls well below the observed dust luminosity (Figure \ref{f6}), for an assumed $0.1~M_{\odot}$ of $^{56}$Ni formed in the explosion.  Furthermore, the observed dust luminosity varies slowly and does not match the radioactive decay rate of $^{56}$Co.  While dust condensation may have occurred in the freely expanding ejecta of SN 2005ip, the available radioactive power cannot produce the observed dust grain luminosities at late times.

\subsubsection{Collisional Heating}
\label{sec_csm}

As the forward shock expands into the circumstellar medium, the hot post-shocked electrons may collisionally heat pre-existing dust grains produced by the progenitor wind.  To heat the dust to the observed temperatures (T $\approx$~1400-1600 K), figure 1 of \citet{dwek08} shows that the required gas densities are high, ($n\ga10^6~{\rm cm^{-3}}$), assuming post-shock gas temperatures on order of $10^7$ K.  The sputtering effects for such high gas densities would significantly limit the contribution of infrared emission from collisionally heated dust throughout the duration of the observations.

A stronger argument against this scenario is the inconsistency in the time lag between the SN and the observed infrared emission.  The SN peak luminosity vaporizes all dust grains within a radius $r_{v}$ that, for a dust emission/absorption efficiency \Qo, is proportional to the supernova's peak luminosity $L_{peak}$ \citep{dwek83b},
\begin{equation}
r_v\propto \left(\frac{L_{peak}}{T_d^5}\right)^{0.5}.
\label{eqn_dustrad}
\end{equation}
For a SN with a peak UV-optical bolometric luminosity of $1~\times~10^{10}~$\lsolar~with an exponential decline rate time-scale of 25 days ($L_t=1.0\times10^{10}e^{-t/25d}~$\lsolar), \citet{dwek83b,dwek85} shows $r_v\approx6\times10^{16}$ cm (0.06 light years) for carbon-rich grains ($T_{evap}=1900~K$) and $3~\times~10^{17}$ cm (0.31 light years) for oxygen-rich grains ($T_{evap}=1500~K$).  While the very early evolution of SN 2005ip is not well known, an early spectrum of SN 2005ip (Modjaz, private communication) and $R$-band photometry \citep{smith08b} indicate an observed luminosity  of $\sim0.2\times10^{10} $~\lsolar~and an exponential time-scale of $\sim$35 days.  Considering the supernova was not detected until a few weeks past maximum, the actual peak is most likely greater by a factor of a few, which is comparable to the supernova peak luminosity used by \citet{dwek83b} as described above.

The size of this dust cavity implies a time lag between the initial explosion and the onset of infrared emission equal to the shock propagation time across the dust free radius.  Using the minimum of the H$\alpha$ absorption line, \citet{modjaz05} measure the SN expansion velocity to be $1.5\times10^9$~\cms~(0.05$c$) at an early age.  This speed refers to the photosphere and there is presumably higher velocity gas, but there is also likely to be deceleration resulting from the interaction.  At this speed the shock must travel for $\sim$450 days before encountering the dust, assuming the dust cavity radius for carbon-rich grains given above, and $\sim$2300 days for the oxygen-rich grains.  Figure \ref{f4}, however, shows a considerably shorter time delay before a detectable infrared excess.

\subsubsection{IR Echo}
\label{sec_echo}

The IR echo model given by \citet{dwek83b} is tempting to explain SN 2005ip because it predicts an extended luminosity plateau, as observed in Figure \ref{f6}b.  Again, the fading photosphere is responsible for the large temperatures and rapidly evolving luminosity over the first 150 days in Figure \ref{f6}b.  Once the photosphere fades sufficiently, the dust temperature is $\sim$1400-1600 K.  This temperature is approximately the dust vaporization temperature, which is consistent with the dust lying just beyond the cavity predicted by the IR echo model.

Assuming the IR echo scenario to be correct, the duration of the $K_s$-band plateau ($\sim$900 days) indicates the dust cavity radius must be $r_{v}> 1$~light year ($\sim 10^{18}$~cm).  Given the relationship in equation (\ref{eqn_dustrad}), this radius would imply a peak luminosity $L\sim1\times10^{11}$ ~\lsolar.  This luminosity is $\sim10$ times larger than the estimate made in \S~\ref{sec_csm}.    Assuming a fairly exponential decay \citep{smith08b}, the total radiated energy from the supernova over the first few weeks is approximately $E\sim10^{51}$ ergs. 

Given the infrared luminosity plateau in Figure \ref{f6}b, the total radiated infrared energy through day \lastday~is approximately $E_{IR}=3\times10^{41}~\ergs \times (8.1\times 10^7 {\rm~s})\approx2.4\times 10^{49}$ ergs.  This energy is only a lower limit because little decline of the infrared flux has been observed.  Furthermore, this luminosity does not account for cooler dust contributions. With the simplified assumption that these grains are emitting as blackbodies, the optical depth is approximated as $\tau\approx\frac{E_{IR}}{E+E_{IR}}\sim 0.02$, which, like the infrared energy, is also a lower limit.  This optical depth is consistent with the ultraviolet/optical observations of the supernova, which do not indicate a large amount of reddening ($B-V=0.2$ on day 461) \citep{immler07}.

The optical depth gives a good estimate of the gas mass loss from the progenitor.  The above optical depth $\tau\sim0.02$ corresponds to a visual extinction $A_V = 1.086\times\tau = 0.02$.  Assuming dust properties comparable to the Milky Way's interstellar medium, the column density is given as $N_H=1.8\times10^{21}\times A_V = 3.9\times10^{19}~\rm{cm}^{-2}$.  Given the inner dust cavity radius derived above, $r_v \sim10^{18}$ cm, the total mass loss is then $M_{loss} \approx N_h \times (4 \pi r_v^2) \times m_{H} \sim0.5~M_{\odot}$.  Assuming a wind velocity of $\sim$10 \kms, the mass loss rate is $\sim2\times 10^{-5}$ \ml, which agrees with measured values of typical SN winds.

Despite the consistency in the above arguments, the required luminosity for this scenario is remarkably high for a SN and is comparable to the most luminous events that have ever been observed \citep{quimby07}.  Even if an eruption prior to the core collapse could have produced the large dust cavity, the same peak luminosity is required to warm the dust to the observed temperatures.  Early time optical observations do not suggest that the SN was unusually bright at maximum \citep[Teamo 2007,][]{smith08b}.  An IR echo, while able to adequately predict an extended luminosity plateau, is not the likely source of the late-time infrared emission in this case.

\subsubsection{Dust Condensation in Post-Shocked Gas}

A plausible source of warm dust in SN 2005ip is dust condensation in the shocked region that results from the collision of the ejecta with the pre-existing circumstellar medium.  If the shocked gas is able to efficiently cool, the high pressure in the region results in gas densities $\ga 10^{10}$ cm$^{-3}$ when the temperature drops to $\sim 10^3$ K.  The shocked region is bounded by a forward shock in the circumstellar gas and a reverse shock in the supernova ejecta.  Radiative cooling is more likely at the reverse shock because of the higher density, lower shock velocity, and the possibility of heavy element enrichment.  Cooling at the forward shock is less likely, but the presence of clumps in the circumstellar medium, as indicated by the relatively narrow lines observed in SNe IIn, may allow radiative shocks.

While we cannot provide any direct spectroscopic evidence at this time to support dust condensation in a particular shock front, the available photometry is consistent with dust condensation in the shocked region.  The observed dust temperatures and luminosities (both infrared and X-ray) are similar to SNe 1998S, 1995N, and 2006jc.  Like SN 2005ip, these three supernovae have associated X-rays \citep{fox00,pooley02,immler08}, which are evidence for interaction of the supernova shock with the circumstellar gas.  These results indicate that the source of the infrared emission is most likely associated with the shock interaction region.  While SN 1995N was the only one of these supernovae not previously associated with dust condensation in the reverse shock, we consider the possibility below in \S \ref{sec_conclusion}.  

The dust temperature ($T\approx1400-1600~K$) on day $\sim200$ is also consistent with the condensation temperature of carbon-rich grains.  Assuming the dust shell emits as a single, spherical blackbody ($\tau>1$), a minimum radius of $\sim10^{16}$~cm is consistent with the observed luminosity.  This radius corresponds to a shock expansion velocity of $\sim3000-8000~\kms$, where the range is related to the grain composition.  The upper part of this range is quite reasonable.  Unlike the IR echo scenario, this picture is physically coherent and does not require an unreasonably high luminosity. 

The dust equilibrium temperature is ultimately determined by a balance of heating and cooling.  In this case, the dust heating mechanism is likely radiative shock emission.  While the observed X-ray luminosity on day 466 is only $1.6\times 10^{40}$~\ergs \citep{immler07}, the corresponding optical luminosity is $\sim2\times10^{41}$~\ergs, which is deduced from the $V$ magnitude at this time.  This luminosity is comparable to the observed infrared luminosity ($\tau\approx\frac{E_{IR}}{E+E_{IR}}\sim 0.5$).  The $R$-band luminosity plateau \citep{smith08b} suggests that this visible component persists throughout the duration of our observations and, if absorbed by the newly formed dust, is a significant source of heat for the dust grains.

As the cool, dense shell expands, the radiation field from the shock emission becomes diluted and we expect the dust temperature to vary as $T\propto (L/r^2)^{1/4}$.  Assuming an expansion velocity of $\sim5000~\kms$, by day $\sim800$ the dust shell radius would have expanded by about a factor of 3 and the temperature would have dropped from $\sim1500$~K to $\sim850$~K.  Figure \ref{f6}b shows a decreasing $T$ with time, although the decline appears to be slightly weaker than expected.  This discrepancy is most likely related to the accuracy of our assumptions involving expansion velocity and grain composition.

\section{Discussion and Conclusions}
\label{sec_conclusion}

This paper presented near-infrared observations of SN 2005ip for the first \lastday~days following detection.  A large $K_s$-band luminosity persists even as the supernova's $J$-band luminosity falls.  Among a variety of potential mechanisms, dust condensation in the cool, dense shell downstream from the reverse shock is the likely source of the observed infrared emission.  

We are able to rule out other mechanisms for late-time near-infrared emission.  While dust condensation in the supernova ejecta is a likely possibility, the duration of the observed near-infrared light curve is inconsistent with the quickly declining radioactive heating source.  Shock/mechanical heating of pre-existing dust grains from prior mass loss is not possible because the near infrared excess appears quickly ($<$100 days), providing insufficient time for the expanding ejecta to cross the dust-free cavity formed by the supernova.  An IR echo successfully explains the fairly uniform luminosity over the period of observation.  The duration of the infrared excess, however, implies a cavity size that requires a peak supernova luminosity that is much larger than observed for SN 2005ip and is comparable to the most luminous supernova ever observed.  

On the other hand, there is mounting evidence to support dust condensation in the shock interaction region.  SN 2005ip shares many properties with SNe 1998S and 2006jc, for which spectra reveal direct evidence of dust condensation in the reverse shock \citep{pozzo04,smith08b}.  While we currently have no such spectra for SN 2005ip, the observed infrared luminosities and temperatures are consistent with dust forming in shocked, cooled gas.  The observed X-rays indicate the infrared emission is most likely associated with the shock interaction region.  Furthermore, the initial dust temperature of $\sim1400-1600$~K is consistent with the dust condensation temperature of carbon-rich grains, which we expect to find in the ejecta encountering the reverse shock.  The temperature evolution is also somewhat consistent with the decline predicted by an expanding cool, dense shell and diluted radiation field.  Unlike the IR echo, this scenario is physically plausible and a large initial luminosity is not required.

Of course, near-infrared photometry is only a small part of a much larger picture.  Infrared spectra can better constrain the dust temperature, infrared flux, and emission/absorption efficiency than can be obtained with near-infrared photometry alone.  Mid-infrared spectra also hold the prospects of revealing the mineralogy of the heated dust, distinguishing between carbon-rich (graphite) and oxygen-rich (silicate) grains.  Based on the dust composition and mass loss history, the nature of the progenitor system may be deduced.  Optical spectra are also of interest in that they can potentially reveal more evidence for dust formation and heating via circumstellar interaction \citep[e.g.][]{fransson02}

The overall similarity between SN 1995N and SN 2005ip suggests that SN 1995N be considered in the context of the dust condensation model, as opposed to the echo model favored by \citet{gerardy02}.  Indeed, \citet{pastorello05} show that the late-time optical and infrared luminosities of SN 1995N are comparable, which suggests a correlation between the ejecta/circumstellar medium interaction and the warm dust.  The IR echo model also predicts a optical light echo from the scattered light, as well as an infrared echo produced by the absorption of the supernova peak luminosity \citep{chevalier86}.  Yet detailed spectroscopic observations of SN 1995N suggest the late visible emission is due to circumstellar interaction and not an echo \citep{fransson02}.  These spectra reveal evidence for absorption in the diminution of the red sides of emission lines compared to the blue sides, as would be expected with dust formation.  Furthermore, the authors present evidence for an increased helium abundance, CNO-burning products, and Ly$\alpha$-pumped fluorescence of Fe II lines, all of which suggest a complex interaction between the ejecta and a pre-existing, dense circumstellar medium.  The dust temperature of SN 1995N on day 730, $700-800$~K, is slightly smaller than for SN 2005ip.  While this difference may be not be statistically significant, it is possible that SN 1995N had either a significantly higher expansion velocity or a different grain composition.

Our discussion of SN 1995N and the data for SN 2005ip indicate that Type IIn SNe may be sources of dust in our universe.  This association would explain the consistent late-time infrared emission observed in several Type IIn events \citep[e.g.][]{pastorello02,gerardy02,pozzo04}.  The rarity of Type IIn events, which represent only $\sim$2-3\% of all core-collapse SNe \citep{galyam07}, makes collecting complete light curves difficult.  Nonetheless, an interesting study would include follow-up observations of all Type IIn events over the past several years to determine which events are still bright in the infrared.  We also reiterate the call by \citet{gerardy02} for simultaneous UV and X-Ray observations to better constrain $\tau$ and the amount of shock radiation being reprocessed by the dust grains.

\acknowledgments

The authors would like to thank the NSF for the AAG-0607737.  Ori Fox would also like to thank the NASA Graduate Student Researchers Program for their financial support.  RAC was supported in part by NSF grant AST-0807727.  The infrared imaging camera at UVa's Fan Mountain Observatory was supported by NSF grant AST-0352934 and the F.H. Levinson Fund of the Peninsula Community Foundation.  The authors thank the anonymous referee and Seppo Mattila for their useful comments on the manuscript.  Ori would also like to thank Eli Dwek for his generous insight.

\clearpage
\bibliographystyle{apj}
\bibliography{references}

\end{document}